\documentclass[reprint,superscriptaddress,preprintnumbers,nofootinbib,amsmath,amssymb,aps]{revtex4-2}

\usepackage{graphicx}
\usepackage{dcolumn}
\usepackage{bm}

\usepackage{amssymb}    
\usepackage{color}         
\usepackage{graphicx}     
\usepackage{mathrsfs} 
\usepackage{hyperref} 
\hypersetup{colorlinks=true,linkcolor=blue,citecolor=blue}
\usepackage{ upgreek } 
\usepackage{color,xcolor,fancybox,epsf,rotating,colordvi}

\newcommand{\foods}{DarkSHINE}

\newcommand{\fooeot}{EOTs}

\begin{document}

\preprint{The 32nd International Symposium on Lepton Photon Interactions at High Energies, \\ Madison, Wisconsin, USA, August 25-29, 2025}

\title{DarkSHINE: Search for Light Dark Matter at the SHINE Facility in Shanghai}
\author{Haijun Yang (on behalf of the DarkSHINE Working Group)}
\email[Corresponding author:]{haijun.yang@sjtu.edu.cn} 
\affiliation{School of Physics and Astronomy, Shanghai Jiao Tong University, \\  800 Dongchuan Road, Shanghai 200240,  China}
\affiliation{Tsung-Dao Lee Institute, Shanghai Jiao Tong University, \\ 1 Lisuo Road, Shanghai 201210, China}
\affiliation{State Key Laboratory of Dark Matter Physics, \\Key Laboratory for Particle Astrophysics and Cosmology (MOE), 
\\Shanghai Key Laboratory for Particle Physics and Cosmology (SKLPPC)}

\date{\today}

\begin{abstract}
 DarkSHINE is an electron-fixed-target experiment under proposal that aims to probe light dark matter in the MeV–GeV mass range via the invisible decay of dark photons (A'), leveraging the High repetition-rate 8 GeV electron beam from the Shanghai High repetition-rate XFEL and Extreme Light Facility (SHINE). 
 This paper presents the core detector design of the experiment, R\&D, the simulation framework, and the prospects of the physics. The detector system integrates an AC-coupled Low Gain Avalanche Diode (AC-LGAD) silicon tracker, a LYSO crystal electromagnetic calorimeter (ECAL), and a scintillator-based hadronic calorimeter (HCAL), all optimized for SHINE’s high-radiation, high-rate environment. The prototype tests at DESY and CERN have validated key performance metrics, including a spatial resolution of 6.5-8.2 $\mu m$ for AC-LGAD silicon strip sensor, an ECAL energy resolution of  $1.8\%/\sqrt{E(GeV)} \bigoplus 0.66\%$. Based on MC simulations and $9 \times 10^{14}$ electrons-on-target (EOT), the DarkSHINE experiment is expected to rule out most of the sensitive regions predicted by popular dark photon models. 
 \end{abstract}

\maketitle
\newpage


\section{\label{sec:Introduction}Introduction}

Based on the results from cosmology and astronomy in the past few decades~\cite{PhysRevD.76.083012}, there is strong evidence that dark matter (DM) exists in the universe. DM is about five times more than ordinary matter~\cite{Planck:2013jfk}, however, its fundamental nature remains one of the most pressing mysteries in modern physics. 
In the last decades, searching for DM candidates using alternative approaches to astronomy observatories have been launched, such as space experiments involving DAMPE~\cite{Kyratzis:2022gvg} and AMS~\cite{Giovacchini:2020vxz}; collider experiments involving LHC~\cite{Behr:2022tyz}, BELLE-II~\cite{Laurenza:2022rjm}, and BES-III~\cite{Prasad:2019ris}; and underground experiments involving PandaX~\cite{PandaX-4T:2021bab}, CDEX~\cite{CDEX:2022fig}, LUX~\cite{LUX:2022vee} and Xenon~\cite{XENON:2021myl}. A large parameter space of DM in the GeV-TeV mass range has been ruled out~\cite{Billard:2021uyg}.

Although Weakly Interacting Massive Particles (WIMPs) in the GeV–TeV range have long dominated search efforts, mounting theoretical and experimental evidence has shifted focus to light dark matter (LDM) with masses between MeV and GeV, a regime largely unexplored due to limitations of traditional detection methods. 

The dark photon ($A'$), a hypothetical abelian gauge boson, emerges as a vital portal between the standard model (SM) and DM\cite{Fuyuto:2019vfe,Choi:2020dec,Cheng:2021qbl}, in analogy to the ordinary photon of electromagnetism ~\cite{Holdom:1985ag,Foot:1991kb}. Dark photons can couple with photons with a kinetic mixing parameter $\varepsilon$ and then interact with SM particles.  
Several experiments aim to search for dark mediators and LDM particles using fixed-target or collider experiments, including NA64 at CERN~\cite{NA64}, BES\uppercase\expandafter{\romannumeral3} at BEPC\uppercase\expandafter{\romannumeral2} \cite{Zhang:2019wnz}.
More experiments remain in the proposal stage including LDMX~\cite{LDMX:2019gvz,LDMX:Snowmass2022,LDMX:2025}.
DarkLight~\cite{DarkLight}, DarkMESA~\cite{DarkMESA} and DarkQuest~\cite{DarkQuest}, which help to establish the new frontier for accelerator based fixed-target DM search experiments worldwide. 


\section{SHINE Beamline}

The Shanghai high repetition rate XFEL and extreme light facility (SHINE) is under construction and aims to deliver its first electron beam in 2026~\cite{SHINE:2021,SHINE:2017,SHINE:2016}. 
The SHINE facility is designed to deliver an electron beam with an energy of 8 GeV, operating at 1 MHz with 100 pC per electron bunch. This beam is dedicated to XFEL light source generation. 

\begin{figure}[!htbp]
    \centering
    \includegraphics[width=0.95\linewidth]{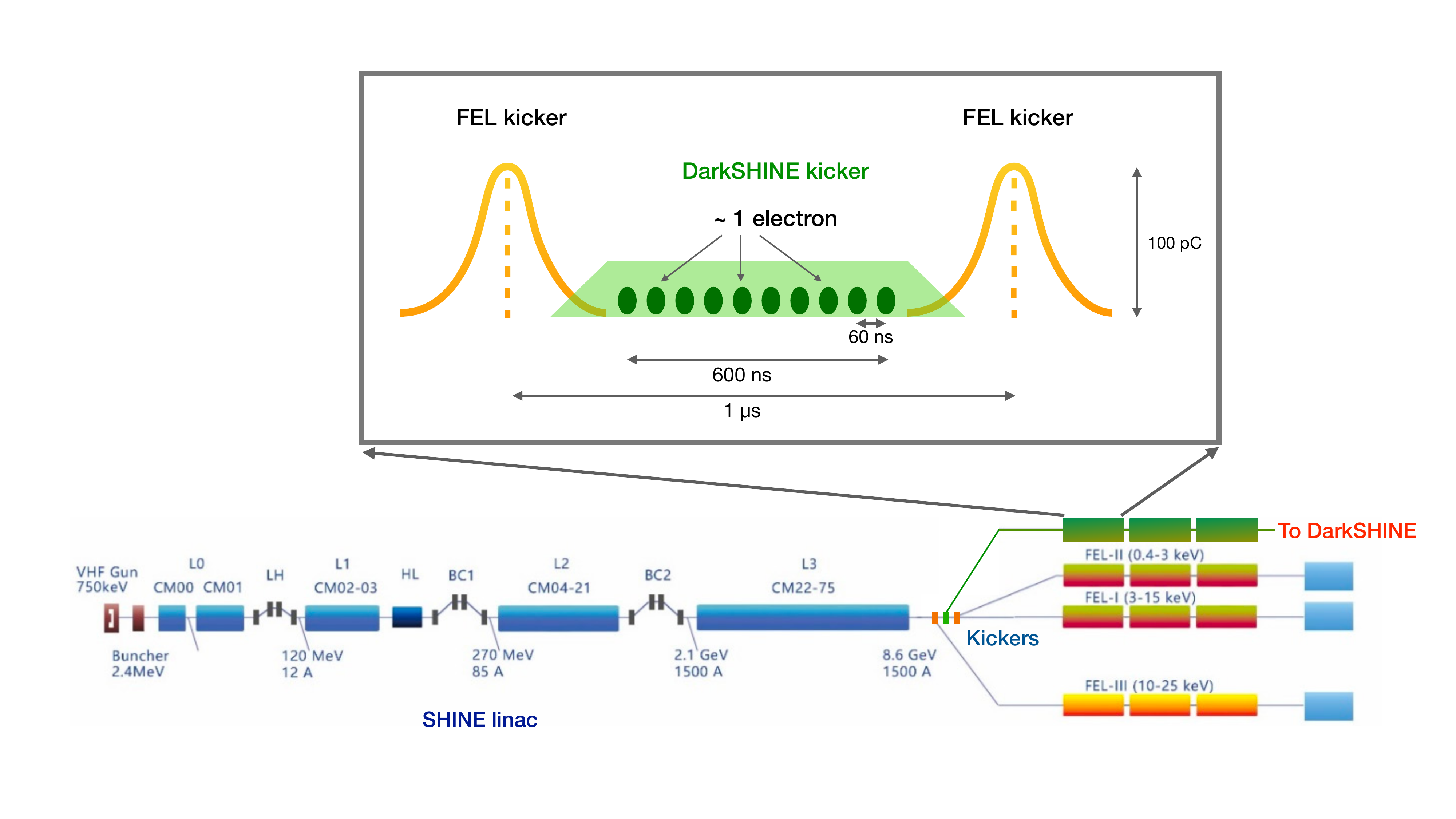}
    \caption{The conceptual design of the FEL kicker, DarkSHINE kicker and beamline.}
    \label{fig:DarkSHINE-kicker}
\end{figure}

For the DarkSHINE experiment, the SHINE linac requires the development of a dedicated single-electron beam (one electron per bunch). This single-electron beam is generated by a low-power laser and then injected into the ten empty buckets between two adjacent FEL bunches via a kicker system, as illustrated in FIG.~\ref{fig:DarkSHINE-kicker}. This configuration yields a 10 MHz repetition rate for the DarkSHINE-dedicated single-electron beam.

DarkSHINE~\cite{DarkSHINE:SCPMA2023,DarkSHINE:LP2023}, a new initiative, proposes to conduct searches for dark mediators and LDM via an electron-on-target (EoT) experiment. This approach leverages the high-repetition-rate single-electron beam that will be provided by the SHINE facility.

DarkSHINE’s benchmark design places special emphasis on probing the invisible decays of the dark mediator - the Dark Photon. This vector boson bridges the SM and DM sectors by kinetically mixing with SM photons. In one of the most simplified scenarios, an additional U(1) symmetry is introduced, which predicts a new gauge field (X) and the corresponding Dark Photon vector boson~\cite{Holdom:1985ag,Foot:1991kb}. The unique characteristics of invisible decays drive the conceptual design of the DarkSHINE detector.

\section{DarkSHINE Detector Design}

\begin{figure}[!htbp]
    \centering
    \includegraphics[width=0.95\linewidth]{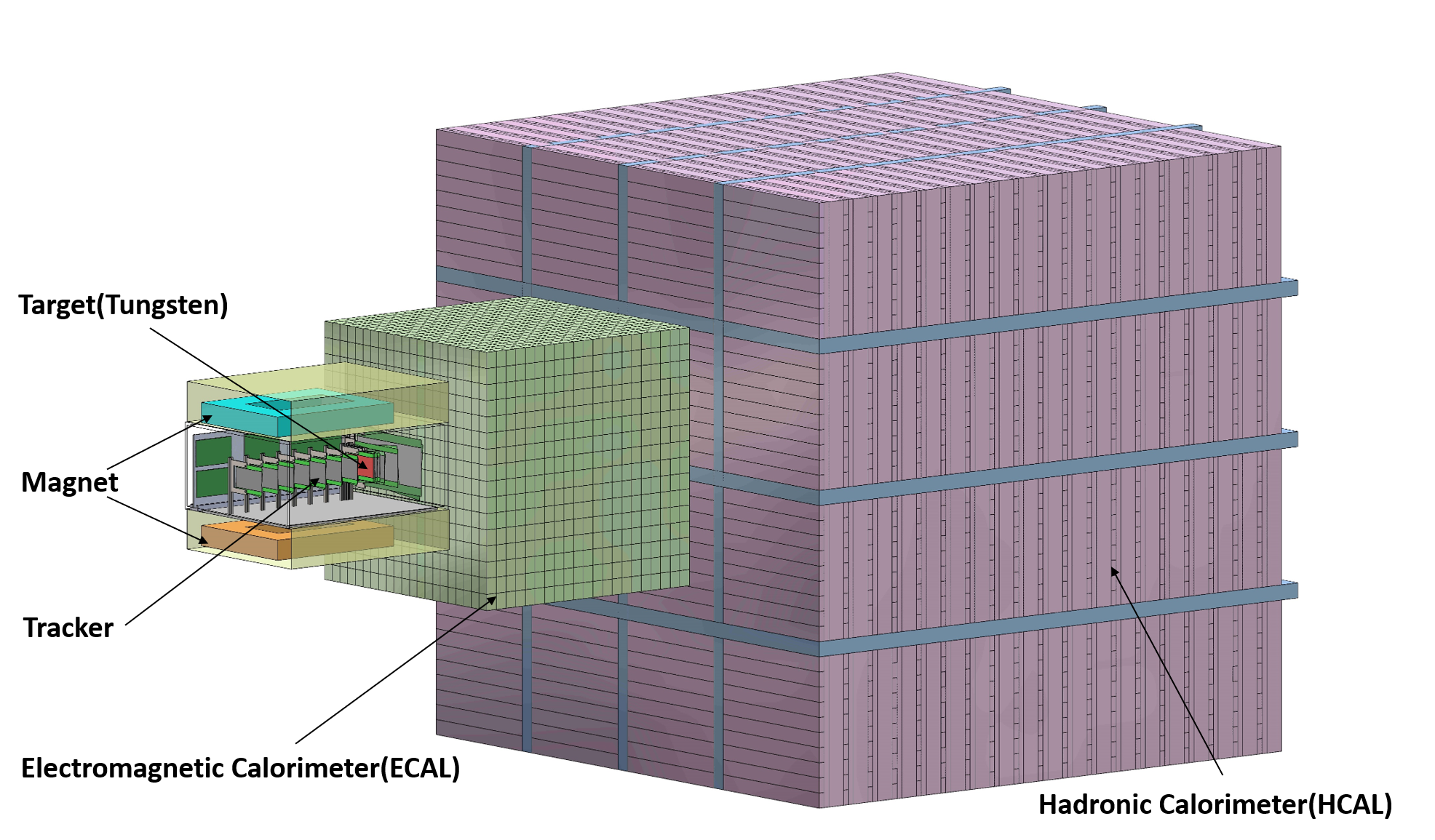}
    \caption{The conceptual design of the DarkSHINE detector is structured as follows: along the electron incident direction (from left to right), a tagging tracker, tungsten target, and recoil tracker are mounted inside a dipole magnet with a 1.5 T magnetic field. These are followed by a crystal electromagnetic calorimeter (ECAL) encircled by a side hadronic calorimeter (sideHCAL), and a large hadronic calorimeter (HCAL) positioned downstream.}
    \label{fig:DarkSHINE-detector}
\end{figure}

The DarkSHINE detector system is engineered to precisely measure recoil electron trajectories and energies while efficiently vetoing SM background processes. Its layout comprises a tungsten target, magnetic field system, AC-LGAD silicon tracker, LYSO ECAL, and scintillator HCAL, supported by high-speed readout electronics—all optimized for the high-rate, high-radiation environment of the SHINE facility. The conceptual design of the DarkSHINE detector is shown in FIG. ~\ref{fig:DarkSHINE-detector}.

\subsection{Beam and Target}

DarkSHINE utilizes an 8 GeV single-electron beam (1 electron per bunch) with 10 MHz repetition extracted from SHINE’s linear accelerator via a dedicated low power laser and kicker system. The beam is directed to a thin tungsten target with thickness of 350 microns (0.1 $X_0$), chosen to balance dark photon production efficiency (via bremsstrahlung and t/s-channel processes) against background generation from beam scattering. A 1.5 T uniform magnetic field surrounds the tracker region to bend electron trajectories, enabling precise momentum measurement of incident and recoil electrons, a critical capability for reconstructing the missing momentum vector associated with invisible dark photon decays.

\subsection{Silicon Tracker}

The tracker system is central to DarkSHINE’s physics goals, as it enables reconstruction of electron trajectories and calculation of missing momentum. It employs AC-coupled Low Gain Avalanche Diode (AC-LGAD) sensors, selected for their fast response, radiation hardness, and high spatial resolution, essential qualities for operation in SHINE’s high-rate environment.

The tracker consists of two functional modules: a seven-layer tagging tracker for incident electrons and a six-layer recoil tracker for scattered electrons, with the tungsten target positioned between them. Each layer comprises AC-LGAD strip sensors with 50 $\mu m$ strip width and 50 $\mu m$ gap, and adjacent layers are rotated by 100 mrad to enable 2D track reconstruction. Track reconstruction is performed using a custom algorithm integrated into the simulation framework, which employs hit clustering and Kalman filtering to achieve high reconstruction efficiency.

\begin{figure}[!htbp]
    \centering
    \includegraphics[width=0.6\linewidth]{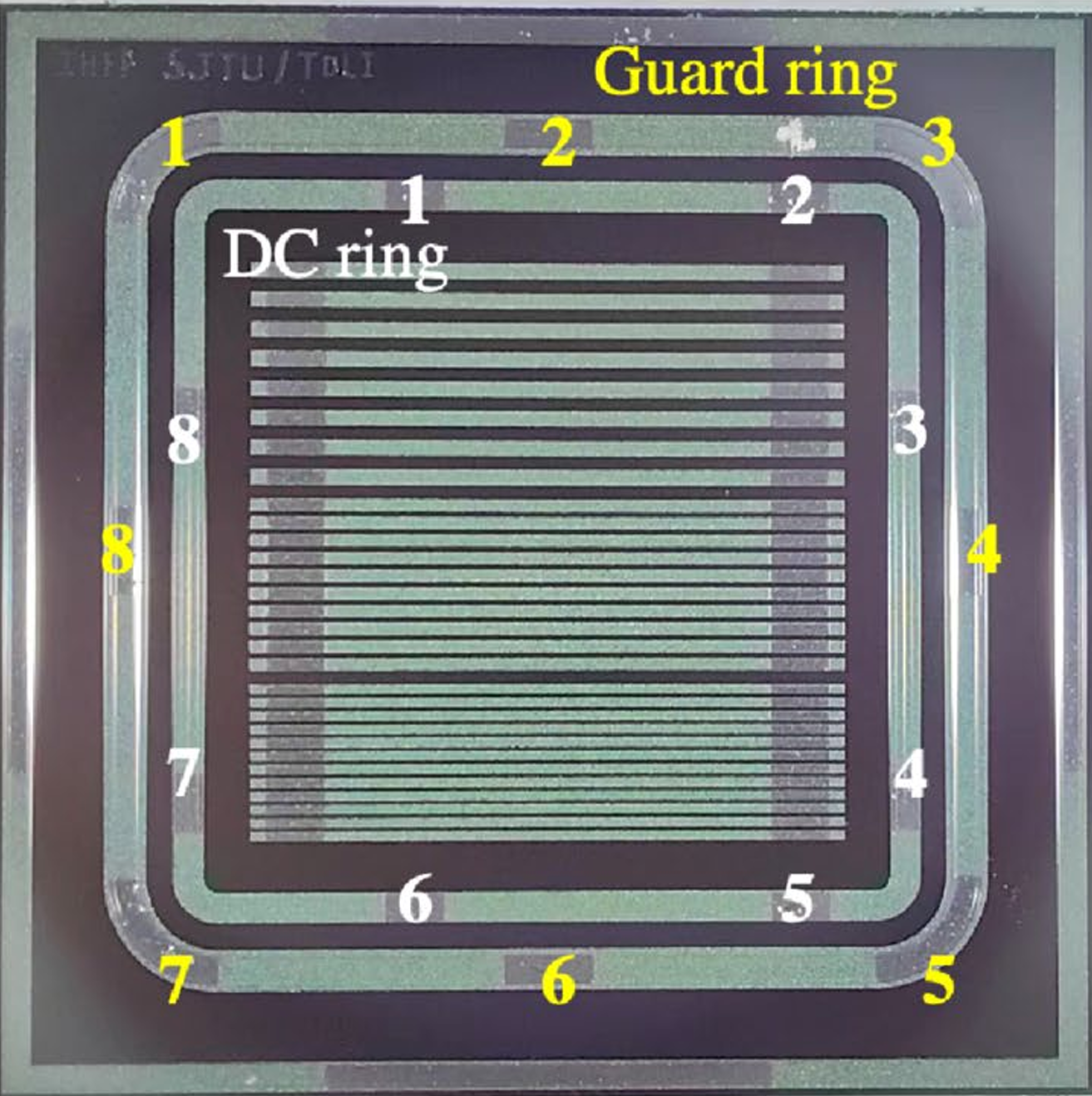}
    \caption{Image of the AC-LGAD silicon strip sensor.}
    \label{fig:DarkSHINE-ACLGAD}
\end{figure}

FIG.~\ref{fig:DarkSHINE-ACLGAD} shows photograph of a fabricated AC-LGAD silicon strip sensor prototype with size of 3638 $\mu m$ $\times$ 3638 $\mu m$. Performance test using infrared laser systems shows spatial resolution of 6.5 to 8.2 $\mu m$ for AC-LGAD sensor with low $n^+$ dose (0.01P) ~\cite{DarkSHINE:AC-LGAD2024}.

\subsection{Electromagnetic Calorimeter}

The ECAL is a homogeneous, full-absorption calorimeter designed to measure the energy of recoil electrons and bremsstrahlung photons with high precision, directly contributing to sensitivity to dark photon signals. It employs LYSO crystal scintillators, chosen for their superior properties: high light yield ($\sim$30,000 ph/MeV) ensuring good signal-to-noise ratio, short decay time ($\sim$40 ns) enabling fast signal processing at 10 MHz beam rates, and excellent radiation hardness.

\begin{figure}[!htbp]
    \centering
    \includegraphics[width=0.7\linewidth]{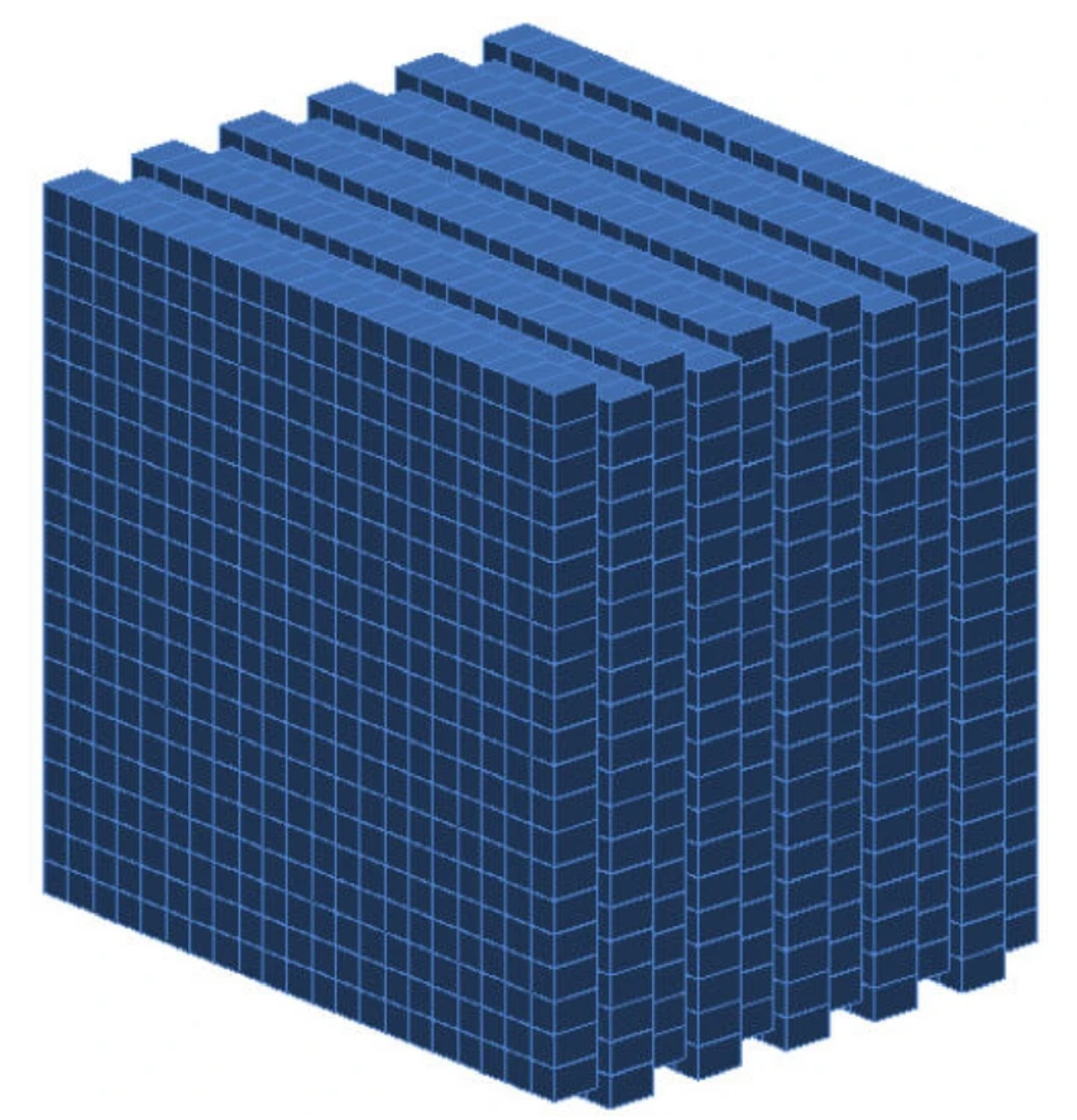}
    \caption{Configuration of the ECAL.}
    \label{fig:DarkSHINE-ECAL}
\end{figure}

FIG.~\ref{fig:DarkSHINE-ECAL} shows the configuration of the ECAL system, a 21×21×11 staggered array of 2.5×2.5×4 $cm^3$ LYSO crystals, resulting in overall dimensions of 52.5×52.5×44 $cm^3$, this geometry balances full absorption of electrons up to 8 GeV with cost and space constraints. Each crystal is coupled to a silicon photomultiplier (SiPM) for light detection, paired with a 10 MHz repetition rate readout electronics system~\cite{DarkSHINE:readout2025}. The electronics features dual-channel high-speed ADCs (14-bit resolution, 1 GSps sampling rate) for waveform digitization, low-noise trans-impedance amplifiers with a double-gain readout scheme (1000× dynamic range), and temperature detection for SiPM gain calibration. 

\begin{figure}[!htbp]
    \centering
    \includegraphics[width=0.7\linewidth]{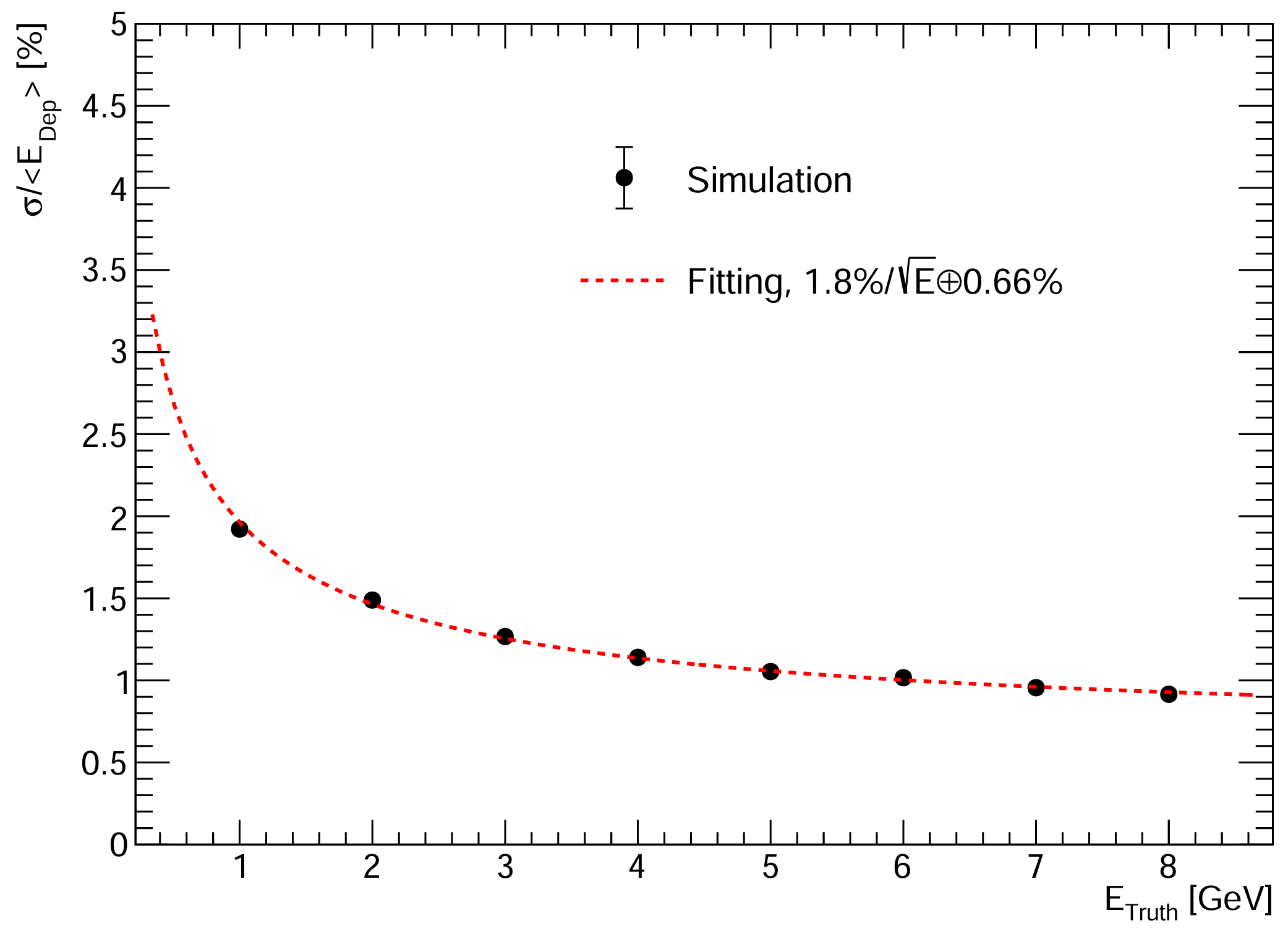}
    \caption{Energy resolution and containment of the ECAL.}
    \label{fig:DarkSHINE-ECAL-res}
\end{figure}

A dedicated digitization model was developed to parameterize the scintillation, SiPM, and ADC behaviors, providing a realistic representation of the detector’s performance. The energy resolution of crystal ECAL is $1.8\%/\sqrt{E(GeV)} \bigoplus 0.66\%$ based on MC simulation as shown in FIG.~\ref{fig:DarkSHINE-ECAL-res} ~\cite{DarkSHINE:ECAL2025}.

Beam tests of ECAL mini-prototype module with 2x2 LYSO crystals have yielded promising results. A 2023 test at DESY’s TB22 beamline, using 1–6 GeV electrons, collected 150,000 events and confirmed the detector’s linear energy response, with signal output scaling consistently with incident energy as predicted by simulations. In 2025, a small prototype was tested at CERN’s PS T9 and SPS H2 beamlines, where 300,000 valid events were collected using electron, muon, and pion beams. These tests validated the ECAL’s performance under mixed-particle environments.

\subsection{Hadronic Calorimeter}

The HCAL’s primary role is to veto muon and hadron backgrounds, particularly for neutrons, which constitute the dominant SM contamination in the dark photon signal region. Its sensitivity to low-energy neutrons and muons is critical for suppressing backgrounds from photon bremsstrahlung and rare meson decays.

The HCAL adopts a sampling calorimeter design, featuring plastic scintillator strips (75 cm length × 5 cm width × 1 cm thickness) encapsulated in carbon sleeves. Wavelength-shifting fibers are embedded for scintillation light collection, with the light detected by SiPMs. Steel absorbers are interleaved with the scintillator layers to enhance hadronic shower development. The detector geometry was optimized via Geant4 simulations, with absorber thicknesses varied across different regions: 10 mm for the first 70 layers and 50 mm for the subsequent 18 layers~\cite{DarkSHINE:HCAL2024}. 

\begin{figure}[!htbp]
    \centering
    \includegraphics[width=0.8\linewidth]{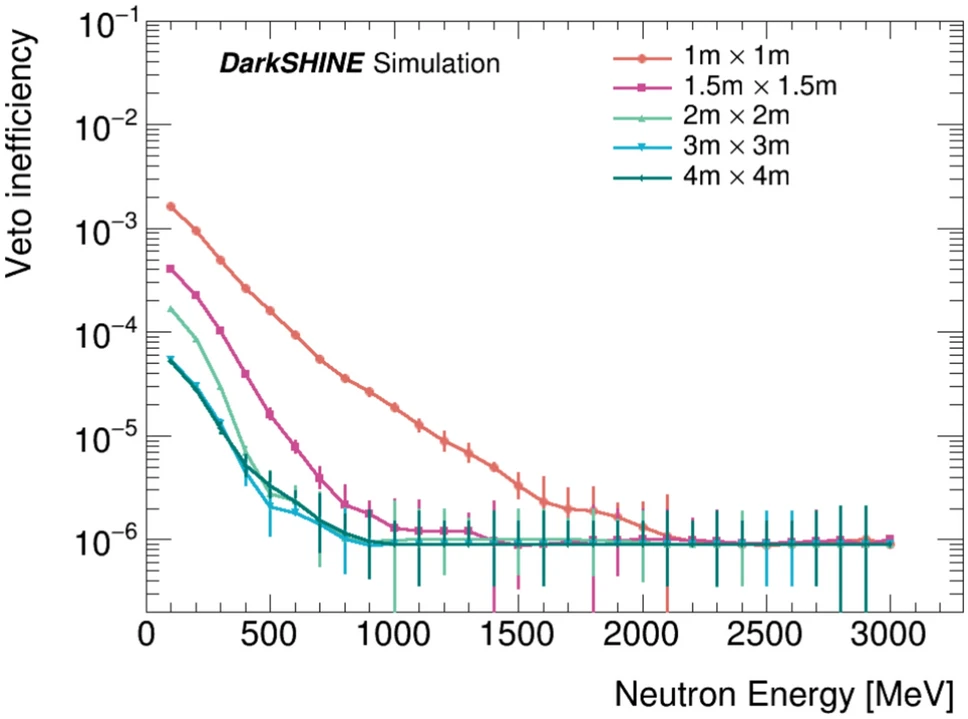}
    \caption{Veto inefficiency as a function of neutron energies with different size of HCAL in x-y plane.}
    \label{fig:DarkSHINE-HCAL-veto}
\end{figure}

This optimization improved veto efficiency for sub-GeV low-energy neutrons while complying with the experimental hall’s load-bearing capacity and budget constraints. 
The HCAL features a 1.5×1.5 m$^2$ active area in the x-y plane and extends 2.5 m along the z-direction (beam direction)—corresponding to over 10 interaction lengths ($\lambda_I$) along the beam path—ensuring efficient detection of hadrons.
Its acceptance aligns with that of the ECAL, ensuring comprehensive coverage of background particles. To further mitigate background contributions from scattered particles—generated by electron-nuclear or photon-nuclear reactions within the calorimeters, a side HCAL is deployed around the ECAL. The side HCAL reduces veto inefficiency by a factor of 3.5. 
For neutrons with energy above 1 GeV, veto inefficiency is minimized to as low as $10^{-6}$, as shown in FIG.~\ref{fig:DarkSHINE-HCAL-veto}.

\section{Simulation and Analysis Framework}

DarkSHINE’s simulation and analysis chain, collectively named DSimu-DAna, is developed to model detector response, optimize performance, and predict sensitivity. Built on Geant4 and ROOT, the framework integrates event generation, detector response modeling, and data analysis to provide a comprehensive tool for experiment design and physics projection.

The DSimu simulation module generates both signal and SM inclusive background events. Dark photon signal events are produced via a custom Monte Carlo generator, encompassing bremsstrahlung, t-channel, and s-channel production mechanisms, while SM background events—including photon bremsstrahlung, electron-nucleus scattering, and meson decays—are generated using Pythia 8. The DSimu includes detailed models of detector response. The DAna analysis module provides tools for event reconstruction, signal selection, and background suppression. 

\section{Sensitivity Projections}

\begin{figure}[!htbp]
    \centering
    \includegraphics[width=0.9\linewidth]{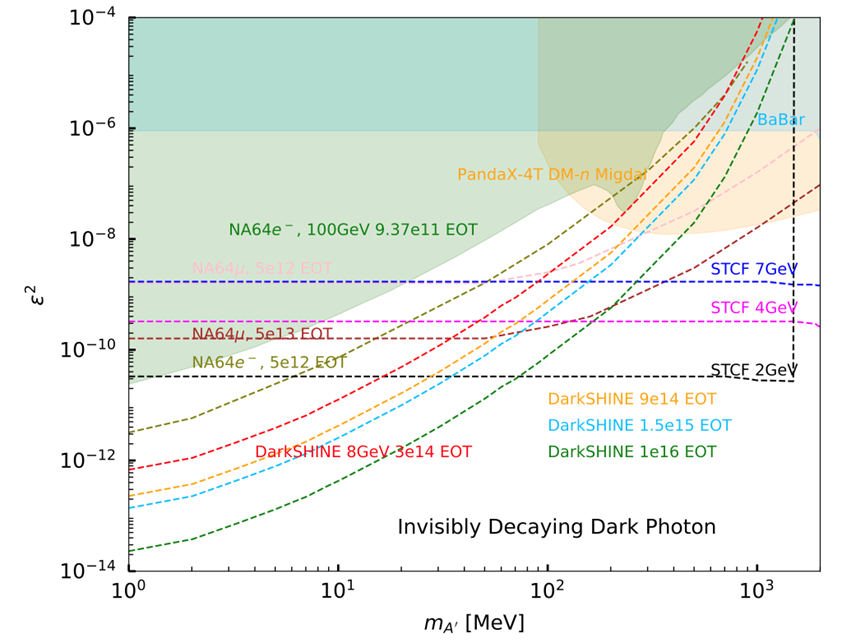}
    \caption{The expected 90\% CL exclusion limits on $\varepsilon^{2}$ as a function of the dark photon mass $m_{A'}$.}
    \label{fig:DarkSHINE-sensitivity}
\end{figure}

DarkSHINE's sensitivity to the dark photon-SM mixing parameter ($\varepsilon^2$) is projected based on detailed MC simulation studies, accounting for signal efficiency, background yield, and exposure time. FIG.~\ref{fig:DarkSHINE-sensitivity} shows current constraints and sensitivity estimates on $\varepsilon^{2}$ as a function of $m_{A'}$, for the expected $3\times 10^{14}$~\fooeot\ (1 year), $9\times 10^{14}$~\fooeot\ (3 years), and $1.5\times 10^{15}$~\fooeot\ (5 years) data collected by the \foods\ experiment. With the phase~II upgrade, we expect $10^{16}$~\fooeot\ and the corresponding limit curve is also shown in the plot. Results from other experiments are shown in the plot for comparison~\cite{BaBar,NA64,Zhang:2019wnz}. The solid curves represent results from operative experiments (BaBar, NA64e), while the dashed curves represent prospective experiment (BESIII, STCF, etc.). Compared with the latest published NA64e result, this study provides improvements by a few orders of magnitude, especially for $m_{A'}$ from MeV to $O(100)$~MeV.

\begin{figure}[!htbp]
    \centering
    \includegraphics[width=0.9\linewidth]{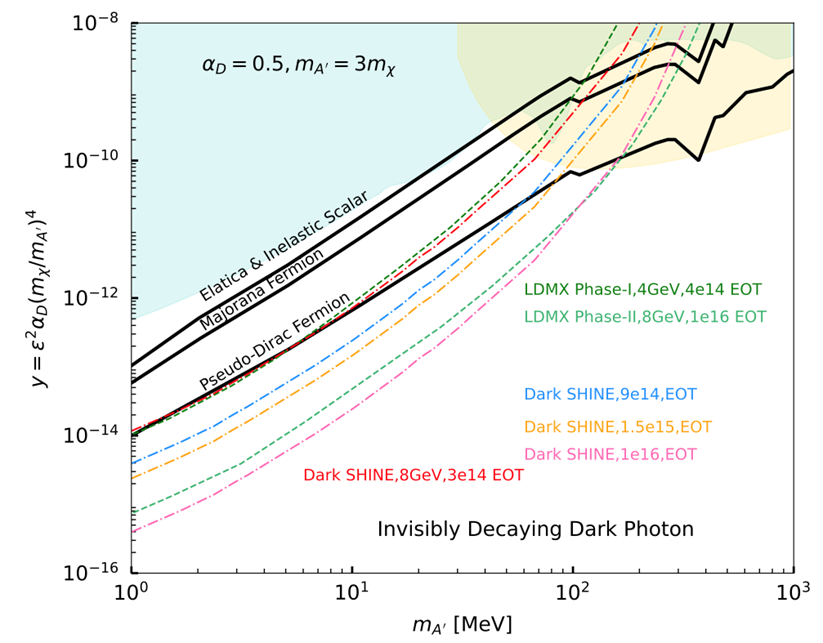}
    \caption{The expected 90\% CL exclusion limit on dimensionless interaction strength $y$ as a function of the DM mass $m_{\chi}$.}
    \label{fig:DarkSHINE-limit}
\end{figure}

In order to explore the thermal relic targets, the projected sensitivity in the dimensionless interaction strength $y=\varepsilon^{2}\alpha_{D}(m_{\chi}/m_{A'})^{4}$ is given as a function of dark matter mass $m_{\chi}$, as shown in FIG.~\ref{fig:DarkSHINE-limit}. It is assumed that the mass of the dark photon $m_{A'}$ is three times as large as the DM mass $m_{\chi}$ and that the coupling constant $\alpha_{D}$ between the dark photon $A'$ and the DM $\chi$ is equal to 0.5. Three benchmark thermal targets are shown in the plot as solid curves: elastic and inelastic scalar, Majorana fermion, and pseudo-Dirac fermion. The filled area are model-dependent constraints provided by several existing experiments and experiment prospective~\cite{NA64,LSND1,LSND2,E137,BaBar,MiniBooNE,directdetection,NA64:2017vtt}. 

The projected sensitivities are estimated for $3\times 10^{14}$, $9\times 10^{14}$, $1.5\times 10^{15}$, and $10^{16}$~\fooeot, respectively, as shown in the colored dashed curves. The results look very promising. With about one year of data taking, the \foods\ would be able to start to probe the existence of the thermal relic DM in the MeV region.
These limits surpass current constraints obtined from experiment like NA64 at CERN by two orders of magnitude in the MeV mass range, filling a critical gap in global LDM searches. 

\section{Conclusion}

DarkSHINE experiment represents a transformative approach to light dark matter search, combining SHINE’s high-repetition-rate electron beam with a state-of-the-art detector system optimized for invisible dark photon decay signatures. 
With projected sensitivity to $\varepsilon^2$ as low as $10^{-12}$ for MeV-scale dark photons, DarkSHINE will surpass current limits and explore uncharted parameter space for LDM. 

The DarkSHINE working group continues to advance R\&D efforts focused on full-scale prototype integration, with plans to assemble a mini-detector prototype integrating tracker, ECAL, and HCAL modules for beam tests. The team is also developing a machine learning-driven trigger system to further reduce background rates.

As China’s first electron-fixed-target dark matter experiment, DarkSHINE will complement global detection efforts and contribute significantly to unraveling the nature of the dark sector.

\begin{acknowledgments}
This work was supported by Ministry of Science and Technology (2023YFA1606904) and National Natural Science Foundation of China (12150006), Shanghai Pilot
Program for Basic Research – Shanghai Jiao Tong University (21TQ1400209), Yangyang Development Fund.
\end{acknowledgments}



\bibliography{ref.bib} 

\end{document}